# Dynamic Response in a Finite Size Composite Multiferroic Thin Film

Zidong Wang (王子东) and Malcolm J. Grimson
*Department of Physics, the University of Auckland, Auckland 1010, New Zealand*
*E-mail address: Zidong.Wang@auckland.ac.nz*

Composite multiferroics, heterostructures of ferromagnetic (FM) and ferroelectric (FE) materials, are characterized by a remarkable magnetoelectric effect at the interface. Previous work has supported the ferromagnetic structure with magnetic spins and the ferroelectric with pseudospins which act as electric dipoles in a microscopic model, coupled with a magnetoelectric interaction [J. Appl. Phys. **118**, 124109 (2015)]. In this work, by solving the stochastic Landau-Lifshitz-Gilbert equation, the electric-field-induced magnetization switching in a twisted boundary condition has been studied, and a behavior of domain wall in the ferromagnetic structure is discussed.

## I. INTRODUCTION

**Composite Multiferroic Materials:**

Recently, the discovery of coupled ferromagnetism and ferroelectricity has strongly revived interest in the field of multiferroism, theoretically [1,2,3,4,5,6] and experimentally [7,8], due to the remarkable effects of the induced magnetization by applying an electric field and the induced polarization by an applied magnetic field [9]. This phenomenon is called magnetoelectric effect. P. Curie firstly discovered this effect in 1894 [10], and it was experimentally confirmed by D. Astrov in 1960 [11]. So far, the mechanism that causes the emergence of the magnetoelectric effect in the ferromagnetic (FM) / ferroelectric (FE) coupled multiferroics is still under discussion. One common origin is the strain-stress coupling [12,13]. Generally, FE materials display the behaviors of piezoelectricity and electrostriction. This provides respectively linear and quadratic shape deformations to an applied field. Similarly, FM materials display piezomagnetism and magnetostriction which can lead an external stress induced magnetic response. Thus, the combination of FM and FE phases generates an enormous magnetoelectric effect [14,15]. Electrostatic screening provides another origin and this has been studied by C. L. Jia et.al [16]. In this paper, we consider a general linear and quadratic magnetoelectric couplings at the FM/FE interface.

**Spin and Pseudospin Models:**

Previous works have discussed a numerical modeling for field-driven composite multiferroics by the spin dynamics approach [1,3,17]. Generally, the technique of the spin dynamics is used to solve the behavior of the magnetic moment due to an effective field in the micromagnetic model [18]. In this model, the magnetic moment is replaced by the spin moment in each individual magnetic spin. The classic Heisenberg model can be used to describe the total free energy stores in this system [19]. FE materials normally contain electric dipoles. In the framework of spin dynamics, the electric dipoles are represented by the pseudospins, with a transverse Ising model to characterize the local energy [20]. The transverse Ising model was conjectured by P. G. de Gennes [21] and R. J. Elliott [22], for the description of order-disorder KDP-type ferroelectrics (e.g. $KH_2PO_4$ and $NaNO_2$). Later the transverse Ising model was also used for displacive type ferroelectrics (e.g. $BaTiO_3$ and $PbTiO_3$) [23]. The novelty here is that we have developed a variable-size pseudospin to compare with the electric dipole in the dielectric materials [24]. A variable-size pseudospin can change its length when an enormous electric field applied, as the separation of an electric dipole.

**Article Outline:**

The aim of present work is to demonstrate the magnetization switching behavior, due to electric field induced polarization with a twisted boundary condition [25]. The numerical result shows the control of the magnetic domain wall structure in the FM lattice [26]. The technical details of the model and the simulation method are introduced in Section II. In Section III, the results are revealed and discussed. Section IV is devoted to the summary and perspectives for future application.

## II. MODEL AND METHOD

**Composite Multiferroic Thin Film:**

In the spirit of a coarse-graining approach, the FM/FE coupled thin film has been considered by a two dimensional heterostructure lattice, with a finite size of $L$ layers in each side and $N$ elements in each layer. The schematic view is in Fig. 1, with the magnetic spins (red arrows) and the electric pseudospins (blue arrows). This lattice is glued by the magnetoelectric coupling at the interface (yellow line) between the last FM layer and the first FE layer. Note that the FM lattice uses twisted boundaries (black arrows) on left- and right-hand sides, in order to show the behavior of the domain wall shifting. The other edges use open (free) boundary condition.

The total energy for the microscopic model can be written as a sum of three main terms in Eq. (1). The first two terms describe the local energies stored in the FM and FE lattices, respectively, and the third term describes the magnetoelectric interaction between the FM and the FE structures.

$$\mathcal{H} = \mathcal{H}_{FM} + \mathcal{H}_{FE} + \mathcal{H}_{ME} \quad (1)$$

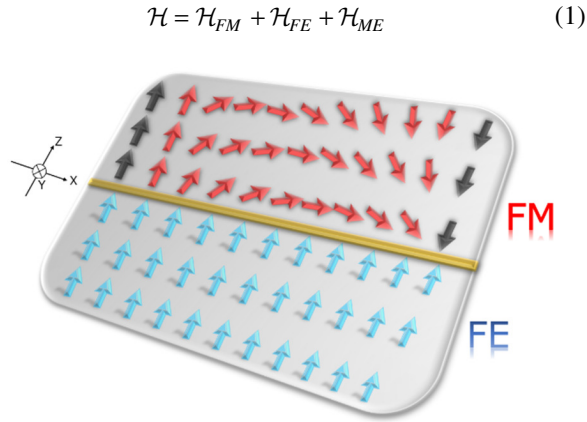

Fig. 1. Schematic illustration of the magnetic spins (red) and the electric pseudospins (blue) in the FM/FE coupled lattice, with magnetoelectric coupling at the interface (yellow). The FM structure with twisted boundaries (black).

**Energy and Dynamics in the FM Lattice:**

In this study, the FM/FE coupled thin film has been regard as a lattice model. In the FM lattice, the magnetic spin $\boldsymbol{S}_{i,j} = (S_{i,j}^x, S_{i,j}^y, S_{i,j}^z)$ is a normalized operator with unit length, i.e. $\|\boldsymbol{S}_{i,j}\| = 1$, $i, j \in [1, 2, 3, \ldots N]$ represents the location of each spin. The Hamiltonian of the magnetic subsystem $\mathcal{H}_{FM}$ is modelled by the classical Heisenberg model, as:

$$\mathcal{H}_{FM} = -J_{FM} \sum_{i,j} \left[ \boldsymbol{S}_{i,j} \cdot \left( \boldsymbol{S}_{i+1,j} + \boldsymbol{S}_{i,j+1} \right) \right] - K_{FM}^z \sum_{i,j} \left( S_{i,j}^z \right)^2 \quad (2)$$

where, the first term represents the nearest-neighbor exchange interaction, and $J_{FM}^* = J_{FM}/k_B T$ is the dimensionless exchange coefficient. The second term represents the uniaxial anisotropy, and $K_{FM}^* = K_{FM}^z/k_B T$ is the dimensionless uniaxial anisotropic coefficient in the $z$-direction.

To describe the time evolution of the spins' response, an differential equation named Landau-Lifshitz-Gilbert equation has been used at atomic level [1,18,27]. In Eq. (3), the Landau-Lifshitz-Gilbert equation predicts the rotation of the magnetization in response to torques. See Movie 1 in online supplementary information in Ref. 28.

$$\frac{\partial \boldsymbol{S}_{i,j}}{\partial t} = -\gamma_{FM} \left[ \boldsymbol{S}_{i,j} \times \boldsymbol{H}_{S_{i,j}}^{eff} \right] - \lambda_{FM} \left[ \boldsymbol{S}_{i,j} \times \left( \boldsymbol{S}_{i,j} \times \boldsymbol{H}_{S_{i,j}}^{eff} \right) \right] \quad (3)$$

where, the gyromagnetic ratio $\gamma_{FM}$ relates the magnetic spin to its angular momentum. $\lambda_{FM}$ denotes the phenomenological Gilbert damping terms in the FM structures. The magnetic effective field $\boldsymbol{H}_{S_{i,j}}^{eff}$ in Eq. (3), is the derivative of the system Hamiltonian of Eq. (2) with respect to the magnitudes of the magnetic spin in each direction, as

$$\boldsymbol{H}_{S_{i,j}}^{eff} = -\frac{\delta \mathcal{H}_{FM}}{\delta \boldsymbol{S}_{i,j}} = \begin{Bmatrix} J_{FM} \left( S_{i+1,j}^x + S_{i,j+1}^x \right) + H_{S_{i,j}^x}^{stoch} \\ J_{FM} \left( S_{i+1,j}^y + S_{i,j+1}^y \right) + H_{S_{i,j}^y}^{stoch} \\ J_{FM} \left( S_{i+1,j}^z + S_{i,j+1}^z \right) + 2K_{FM}^z S_{i,j}^z + H_{S_{i,j}^z}^{stoch} \end{Bmatrix} \quad (4)$$

where $H_{S_{i,j}^{x,y,z}}^{stoch}$ characterizes a stochastic field on the $x$-, $y$- and $z$-components, individually.

**Energy and Dynamics in the FE Thin Film:**

In order to deliver the electric dipole moment into the spin system, here we follow R. J. Elliott and A. P. Young [22], use a pseudospin model with the electric pseudospin $\boldsymbol{P}_{i,j} = (P_{i,j}^z, P_{i,j}^z, P_{i,j}^z)$. $i, j \in [1, 2, 3, \ldots N]$ represent the location of each pseudospin. The Hamiltonian of the pseudospins is depicted by the transverse Ising model $\mathcal{H}_{FE}$ is defined in Eq. (5) with local and external energies.

$$\mathcal{H}_{FE} = -J_{FE} \sum_{i,j} \left[ P_{i,j}^z \left( P_{i+1,j}^z + P_{i,j+1}^z \right) \right] - \Omega_{FE}^x \sum_{i,j} P_{i,j}^x - E_{ext}^z(t) \sum_{i,j} P_{i,j}^z \quad (5)$$

where $J_{FE}^* = J_{FE}/k_B T$ is the dimensionless nearest-neighbor exchange interaction couplings between electric pseudospins. Second term stands for the transverse energy, where $\Omega_{FE}^x$ is the transverse field along the local $x$-axis, which is perpendicular to the Ising $z$-direction in the FE structure [29,30]. Third term is the external energy caused by an applied time-dependent electric field $E_{ext}^*(t) = \epsilon_0 \chi_e E_{ext}^z(t)/k_B T$ in the $z$-direction, where $\epsilon_0$ the electric permittivity of is free space, and $\chi_e$ is the susceptibility. Generally, the polarization $P$ is proportional to the external electric field $E_{ext}$ in the dielectric materials [31]:

$$P = \epsilon_0 \chi_e E_{ext} \quad (6)$$

Here, we develop the size of the electric pseudospin is varied with its effective field $\boldsymbol{H}_{P_{i,j}}^{eff}$ (see Eq. (9) and Ref. 24):

$$\|\boldsymbol{P}_{i,j}\| = \epsilon_0 \aleph_e \|\boldsymbol{H}_{P_{i,j}}^{eff}\| \quad (7)$$

where $\aleph_e$ is the dimensionless pseudoscalar susceptibility.

To solve the time evolution of the electric response in the FE structure we use a spin dynamic method. The limitation of this technique is that the electric dipole moment, which is a measurement of the $z$-component separation of positive and negative charges, is a scalar. Thus, the time evolution of the pseudospins in the FE lattice are expected to perform a precession free trajectory [1,3,24], following

$$\frac{\partial \boldsymbol{P}_{i,j}}{\partial t} = -\lambda_{FE}\left[\boldsymbol{P}_{i,j} \times \left(\boldsymbol{P}_{i,j} \times \boldsymbol{H}_{P_{i,j}}^{eff}\right)\right] \quad (8)$$

where $\lambda_{FE}$ denote the phenomenological Gilbert damping terms in the FE lattice. This is shown in Movie 2 in online supplementary information in Ref. 28. The magnitude of the z-component $P_{i,j}^z$ represents the electric polarization, $P_{i,j}^x$ and $P_{i,j}^y$ are the pseudoscalar polarizations. The electric effective field $\boldsymbol{H}_{P_{i,j}}^{eff}$ in Eq. (8) for the pseudospin, is defined as a functional derivative of Eq. (5).

$$\boldsymbol{H}_{P_{i,j}}^{eff} = -\frac{\delta \mathcal{H}_{FE}}{\delta \boldsymbol{P}_{i,j}} = \begin{Bmatrix} \Omega_{FE}^x \\ 0 \\ J_{FE}\left(P_{i+1,j}^z + P_{i,j+1}^z\right) + E^z(t) + H_{P_{i,j}^z}^{stoch} \end{Bmatrix} \quad (9)$$

where $H_{P_{i,j}^z}^{stoch}$ characterizes the stochastic field. Note that, x- and y-components are the pseudo-components of the pseudospin model and there is no thermal agitation.

**Thermal Effect:**

Thermal effect cannot be neglected. The orientation of the magnetic/electric moments are continuously changed by thermal agitation. We study a simplified Brownian motion by reducing the random forces to a purely random process [1]. This random process is added into the effective field of each magnetic spin or electric pseudospin as a stochastic field $H^{stoch}$, which is a white Gaussian noise, into the dynamics [32]. The probability density function of this stochastic field is given as,

$$\Pr = \frac{1}{\sigma\sqrt{2\pi}}\exp\left[\frac{-\left(H^{stoch}-\mu\right)^2}{2\sigma^2}\right] \quad (10)$$

where $\mu$ is the mean and $\sigma$ is the standard deviation of the Gaussian distribution. Since, the existence of both the magnetization and the electric polarization is required at low temperatures, the standard deviation has been limited to $\sigma = 0.01$, without any bias (i.e. $\mu = 0$). Therefore, both Eqs. (3) and (8) become stochastic Landau-Lifshitz-Gilbert equations.

**Magnetoelectric Interactions:**

The last term in Eq. (1) characterizes the interfacial energy between the last FM layer and the first FE layer, which is described by the dipole-spin interaction Hamiltonian $\mathcal{H}_{ME}$ with the magnetoelectric susceptibility $g_m$ [9,33], as

$$\mathcal{H}_{ME} = -\sum_m \sum_j g_m \left(S_{N,j}^z P_{1,j}^z\right)^m \quad (11)$$

In this work, we only need to account for the low-energy excitations (i.e., piezoelectric/piezomagnetic effects and magnetostrictive/electrostrictive effects) at the interface and so restrict ourselves to the linear $g_1$ and quadratic $g_2$ terms only. Higher order terms have not been studied here, due to their minor relevant effects in the numerical modelling. Thus the interfacial Hamiltonian used in the numerical simulations is

$$\mathcal{H}_{ME} = -\sum_j \left[g_1\left(S_{N,j}^z P_{1,j}^z\right) + g_2\left(S_{N,j}^z P_{1,j}^z\right)^2\right] \quad (12)$$

## III. RESULTS AND DISCUSSION

We implement a numerical simulation for the electric-field-driven multiferroic 16×16 square lattice sample, and a dimensionless parameter site { $J_{FM}^* = J_{FE}^* = 1$, $K_{FM}^* = 0.1$, $\Omega_{FE}^* = 0.1$, $g_1^* = g_2^* = 1$, $\gamma_{FM}^* = 1$ and $\lambda_{FM}^* = \lambda_{FE}^* = 0.1$ }, with twisted boundary conditions in the FM structure, and free boundary conditions for the rest. A rectangular electric field is applied in the local z-axis, with a dimensionless magnitude of $E_0^* = 10$. The simulation proceeds by a fourth order Runge-Kutta method.

The magnetization switching due to electric field induced polarization is shown in Fig. 2. In this figure, the dynamical progress of the mean magnetization $\langle S_z \rangle$ (red curve) and the mean polarization $\langle P_z \rangle$ (blue curve), i.e. the average magnitudes of the z-component in spins and pseudospins, to the normalized driven field $\langle E_{ext} \rangle$ (black dashed-curve). The electric polarization responds to flip over the direction immediately, due to its direct coupling to the electric field. The magnetization then catches up, the switching energy being transferred across the interface from the electric pseudospins by the magnetoelectric effect. The size of the electric pseudospins is limited by the pseudoscalar susceptibility $\aleph_e$. We find $\aleph_e = \left[\left(E_{ext}^z(t)\right)^2 + \left(\Omega_{FE}^x\right)^2\right]^{-1}$ is the most appropriate for the size of electric pseudospin with the unit size of magnetic spin. Hence in Fig. 2, the mean polarization of the electric pseudospin is in the linear dependence region.

In order to observe finer detail, we take a distinct region, dimensionless time $t^* \in [900, 1900]$ shown by the green box in Fig. 2, and study the behavior of the domain wall in the driven part in Fig. 3. The six snapshots show the instantaneous states of the magnetic spins and electric pseudospins in the lattice in Figs. 3(b)-(g). Their relevant time locations are shown in Fig. 3(a) with pink and green symbols, such as "Δ", "∗", "O", "+", "∇" and "X". The multiple colors in Figs. 3(b)-(g) characterize the magnitudes of the spins and pseudospins in the local z-direction. The black line at middle in each panel, represents the interface that divides the FE (front) and FM (behind) structures. In Fig. 3(b) the sequence starts from $t^*("\Delta")=945$. At $t^*("*")=950$, Fig. 3(c) shows the electric pseudospins reorienting. A short time later, $t^*("O")=955$ in Fig. 3(d), the magnetic spins start reorientation close to the interfacial layer. At $t^*("+")=990$, Fig. 3(e), the magnetic spins in the interfacial layer have flipped over and domain wall motion begins to diffuse to the further layers. The domain wall changes its shape when it shifts rightward. Because of the different speeds of the wall propagation in each layer in the FM structure. At $t^*("\nabla")=1165$ in Fig. 3(f), the domain wall shows a more linear-like shape away from the interface. Much later, $t^*("X")=1860$ in Fig. 3(g), the whole domain wall has attained an equilibrium state in the constant driving field that is determined by the twisted boundary condition.

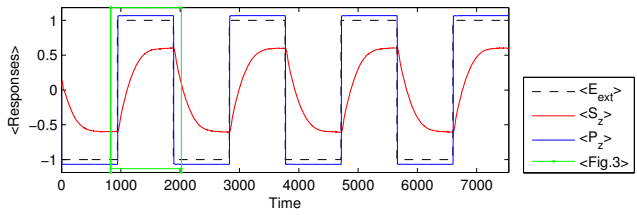

Fig. 2. The dynamic mean *z*-component magnitudes of the magnetic spins (red) and electric pseudospins (blue) are driven by a normalized rectangular electric field (black dashed).

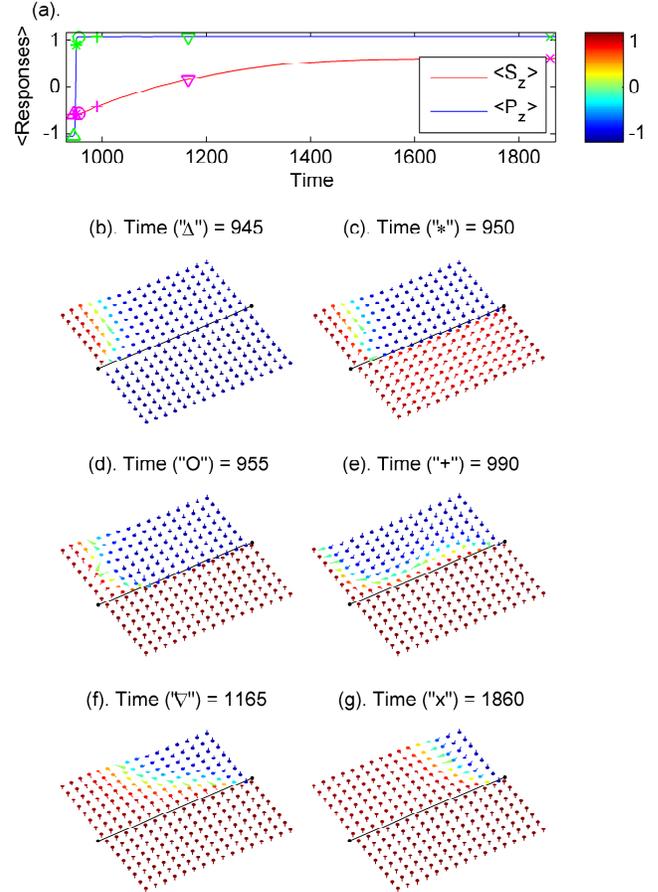

Fig. 3. (a) A closer inspection from Fig. 2 in a distinct region. (b)-(g) Six snapshots (Symbols " $\Delta$ ", " $*$ ", " O ", " + ", " $\nabla$ " and " X " in panel (a)) represent the phase states of the magnetic spins and electric pseudospins in the FM/FE coupled lattice at each particle time. The electric pseudospins are at front of the black line, and behind of the black line are the magnetic spins. The color scale represents the magnitude of the *z*-component. See Movies 3 and 4 in online supplementary information in Ref. 28.

## IV. CONCLUSIONS

The spin dynamics approach allows us to solve the field-driven multiferroic system in a microscopic model. In this paper, the magnetic spin and electric pseudospin have been defined in a two dimensional lattice. We proposed them to investigate the electric-driven-magnetization switching property. In particular, using the twisted boundary condition, the behavior of the magnetic domain wall can be extracted. This analysis can facilitate designing composite multiferroic materials for relevant technological purposes, such as the control of the shifting of the magnetic domain wall via electric field can be applied to develop low-energy consuming electronic devices. Also, the manipulation of the electric properties by an external magnetic field also can be observed in this model.

## ACKNOWLEDGEMENTS

Wang Zidong gratefully acknowledges Wang Feng (王峰) and Zhao Wen Xia (赵雯霞) for financial support.